\documentclass[aps,twocolumn,tightenlines]{revtex4}
\usepackage[dvipdf]{graphicx}
\usepackage{color}
\usepackage{amssymb,amsmath}
\usepackage[latin1]{inputenc}

\begin{document}
\title{Optical pumping of a lithium atomic beam for atom interferometry}

\author{J.~Gillot, A.~Gauguet, M.~B\"uchner, and J.~Vigu\'e}
\address{ Laboratoire Collisions Agr\'egats R\'eactivit\'e -IRSAMC
\\Universit\'e de Toulouse-UPS and CNRS UMR 5589
 118, Route de Narbonne 31062 Toulouse Cedex, France
\\ e-mail:~{\tt jacques.vigue@irsamc.ups-tlse.fr}}

\date{\today}

\begin{abstract}
We apply optical pumping to prepare the lithium beam of our atom interferometer in a single hyperfine-Zeeman sublevel: we use two components of the D1-line for pumping the $^7$Li atoms in a dark state $F,m_F=+2$ (or $-2$) sublevel. The optical pumping efficiency has been characterized by two techniques: state-selective laser atom deflection or magnetic dephasing of the atom interferometer signals. The first technique has not achieved a high sensitivity, because of a limited signal to noise ratio, but magnetic dephasing signals have shown that about $95$\% of the population has been transferred in the aimed sublevel, with similar results for three mean velocities of the atomic beam covering the range $744-1520$ m/s.
\end{abstract}


\maketitle


\section{Introduction}
\label{intro}
In atom optics and interferometry, atom sublevels play a role analogous to polarization states for light. Because of Zeeman effect, in the presence of a magnetic field, the propagation phase varies with the sublevel and this phase dispersion complicates the analysis of experiments such as the one we have done \cite{LepoutrePRL12} to detect the topological He-McKellar- Wilkens phase \cite{HeMcKellarPRA93,WilkensPRL94}. Optical pumping the atomic beam in a single hyperfine-Zeeman sublevel is the ideal way to get rid of this difficulty: this is analogous to the use of a fully polarized light beam in optics. Several experiments have demonstrated the feasibility of an almost perfect pumping of an alkali atomic beam \cite{MastersonPRA93,SchinnJOSAB91}, with up to $97$\% of the atomic population in a single $F,m_F$ sublevel.

We developed a similar optical pumping experiment with our lithium atomic beam produced by supersonic expansion in a rare gas, with all the $^7$Li atoms pumped in their $F=2$, $m_F=+2$ (or $-2$) sublevel. Our experiment has some specificities compare to the other experiments involving the optical pumping of an atomic beam: in particular, our atomic beam is highly collimated and only the collimated beam needs to be pumped. As our beam is injected in an atom interferometer, we have estimated the optical pumping efficiency by measuring the atom interferometer fringe signals in the presence of Zeeman phase shifts.

The paper is organized as follows: section \ref{p1} briefly reviews optical pumping experiments with alkali atomic beams; section \ref{p2} describes our choices concerning the optical pumping experiment; in section \ref{NumSimu}, we present a simplified numerical model of the optical pumping process; in section \ref{ExpResults}, we describe the experimental setup, the experimental results and their analysis.
\section{Brief review of optical pumping experiments of alkali atomic beams}
\label{p1}
In 1950, A. Kastler invented optical pumping : in his seminal paper \cite{KastlerJPR50}, he discussed the optical pumping of a sodium atomic beam, using atomic vapor lamps as the light source, and this experiment was rapidly performed \cite{BrosselJPR52,HawkinsPR53}. This arrangement was sufficient to produce a strong population difference among the ground state sublevels and this difference was used to detect magnetic resonance \cite{BrosselCRAS53}. In the following years, most optical pumping experiments were done in vapor cells, as reviewed by W. Happer \cite{HapperRMP72}.

Several laser beams of different frequencies are required to achieve optical pumping and to characterize the internal state distribution. This is why, the optical pumping of atomic beams developed a lot with the progress of tunable lasers and, especially laser diodes. We will not discuss here optical pumping of cold atom clouds because these experiments usually differ from the optical pumping of a thermal beam and also because there are too many experiments to be quoted here. Here are a short list of papers describing the optical pumping of alkali atomic beams: lithium \cite{BaumAP80,ReichNIMA90}, sodium \cite{DrevesZPA81,HilsAP81,SteelOL83,DrevesPRL83,CusmaPRA83,GouldPRA87,SchinnJOSAB91}, rubidium \cite{FurutaJJAP91,JunPRA98} and cesium \cite{PicqueM77,ArditiJPD78,ArditiM82,WattsOC86,AvilaPRA87,BaumZPD91,MastersonPRA93}. Optical pumping is used for different goals (orientation of the electronic spin, concentration of the largest possible fraction of the atoms in a single $F$ level or in a single $F,m_F$ sublevel) and different applications (atomic clocks, collision studies, parity violation experiments, etc).
\section{Principle of the experiment}
\label{p2}

\subsection{Lithium states and the choice of the pumping transitions}
\label{p21}

We consider here only the most abundant isotope, $^7$Li with a $92.5$\% natural abundance. $^7$Li has a nuclear spin $I=3/2$ and its $^2$S$_{1/2}$ ground state is split in two hyperfine sublevels $F=1$ and $F=2$, with an energy splitting equal to $\Delta E /h= 803.5$ MHz (all energy splittings are expressed in frequency units). We use the first resonance line near $671$ nm for optical pumping. The $^2$P resonance state has two fine structure components (splitting $\approx 10053$ MHz). The  $^2$P$_{1/2}$ is split in two hyperfine levels, $F=1$ and $2$, with a splitting equal to  $92$ MHz while the $^2$P$_{3/2}$ state is split in four hyperfine levels, $F=0$ to $3$, with a total splitting from $F=0$ to $F=3$ equal to $18$ MHz (more details on lithium spectroscopy  in ref. \cite{ArimondoRMP77,SansonettiPRA95,DasJPB08}). The radiative lifetime of the $^2$P state is $\approx 27.1$ ns
\cite{McalexanderPRA96} corresponding to a natural width $\Gamma/(2 \pi) \tau \approx 5.87 $ MHz.

\subsection{Optimum magnetic field value}
\label{p22}
We want to prepare the atoms in the well defined quantum states $F=2$, $m_F=+2$ and $m_F=-2$ according to a quantization axis. Usually, one takes the $z$-axis as the quantization axis, chosen parallel to the magnetic field $\mathbf{B}$. However, as we will refer to previous works on our atom interferometer \cite{MiffreEPJD05} in which the $z$-axis is always along the atomic beam propagation, we keep our choice of axis (see fig. \ref{fig4}) to avoid the risks of confusion. Therefore, the quantization axis is taken parallel to the magnetic field along the $x$-axis. As usual in atomic physics, the polarization of the laser is defined according to this axis: a right-handed circular polarization propagating along the x-axis is $\sigma^{+}$-polarized and $\sigma^{-}$ if it propagates in the other way. In order to be in the ideal setup, we have to cancel the perpendicular components $B_y$ and  $B_z$ of the magnetic field in the pumping volume and produce a $B_x$-component with an optimum value fulfilling opposite requirements:

$B_x$ must be small so that single frequency lasers can excite efficiently all the Zeeman components of the optical pumping transitions i.e. the total splitting of a pumping transition, of the order of $\mu_BB_x$, must be at most comparable to the natural linewidth $\Gamma/(2 \pi)$. This means that  $\mu_BB_x/h \lesssim \Gamma/(2 \pi)$ which is verified if $\left|B_x \right| \lesssim 4 \times 10^{-4}$ T.

$B_x$ must be large in order to minimize the angle between the real field and the laser propagation axis. The minimum $B_x$-value is fixed by the residual values of $B_y$ and $B_z$. With an available gaussmeter, we have obtained $\left|B_{y}\right|\approx \left|B_{z}\right| < 4 \times 10^{-6}$ T and we have chosen $B_x \approx 3.8 \times 10^{-4}$ T so that the angle $\theta = \sqrt{B_y^2 + B_z^2}/\left|B_x\right|$ between the field and the $\mathbf{x}$ axis is smaller than $20$ mrad.

\begin{figure}[h]
\begin{center}
\includegraphics[width= 8 cm]{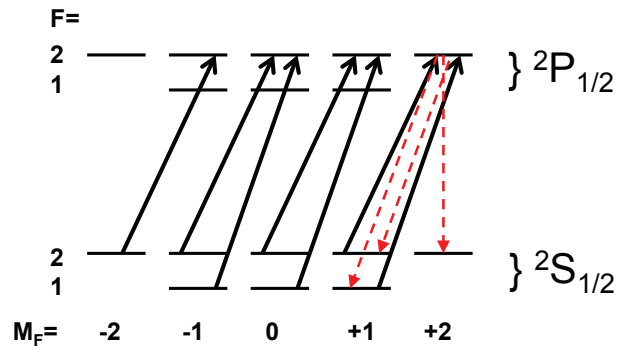}
\caption{(Color on line) The $^2$S$_{1/2}$ and $^2$P$_{1/2}$ levels of lithium atom and the dark state optical pumping scheme.
Two $\sigma^{+}$ polarized laser beams excite the $^2$S$_{1/2}$, $F=1,2 \rightarrow^2$P$_{1/2}$ $F'=2$ transitions.
The dashed lines show the spontaneous emission transitions from the $F'=2,m'_F=2$ sublevel).
\label{fig1}}
\end{center}
\end{figure}

\subsection{Optimum laser parameters}
\label{p23}
 Pumping the lithium atoms in their $F=2, m_F=+2$ sublevel, is achieved with two laser beams: one to empty all the  $F=1, m_F$ sublevels and the other one to transfer the population of the $F=2,m_F$ sublevels in the $F=2, m_F=+2$ sublevel (see fig. \ref{fig1}). This simple view is correct if each laser excites selectively one hyperfine component of the  $^2$S-$^2$P resonance transition: any weak excitation of another hyperfine components would introduce leaks in the pumping cycle. This is possible if the hyperfine structure is well resolved, i.e. if the hyperfine splittings are considerably larger than the natural linewidth. Obviously, the D2-line is not convenient, because the hyperfine structure splittings of the $^2$P$_{3/2}$ state are comparable to the natural linewidth, but the D1-line is more favorable because the hyperfine structure of the $^2$P$_{1/2}$ state is considerably larger than the natural linewidth. As a consequence, we have chosen the D1-line for optical pumping. The laser beam, resonant with the transition $^2$S$_{1/2},F=1 \rightarrow^2$P$_{1/2},F'=2$, empties the $F=1$ level and the laser resonant with the transition $^2$S$_{1/2},F=2 \rightarrow^2$P$_{1/2},F'=2$, pumps the $F=2,m_F$ sublevels in the $F=2,m_F=+2$ sublevel. Both laser beams are circularly polarized ($\sigma^{+}$) in order to induce transitions with $\Delta m_F= m_F' -m_F= +1$ so that the population accumulates in the  $^2$S$_{1/2},F=2, m_F=+2$ sublevel, which is the only dark state (i.e. a state uncoupled to the lasers) if the laser circular polarization is perfect. Figure \ref{fig1} summarizes our optical pumping scheme. Each optical pumping transition is separated from an unwanted transition by the excited state hyperfine splitting and, for small saturation, the line broadening is small so that the excitation probability of the unwanted transition is very small and the associated leak in the optical pumping process may be neglected.

Finally, we need to produce an atomic beam either in the $m_F=+ 2$ or in the $m_F=-2$ sublevel. Due to adiabatic following, the $m_F$ value is conserved throughout propagation, on an axis which is the local magnetic field. As a consequence, we can switch between these two $m_F$ values simply by reversing the magnetic field in the pumping volume.

Our atomic beam is a supersonic beam of lithium seeded in a noble gas. Its longitudinal velocity distribution is approximated by a Gaussian:
\begin{eqnarray}
\label{a2}
P(v)  = \frac{S_{\|}}{v_m \sqrt{\pi}} \exp\left[-\left(\frac{(v-v_m)S_{\|}}{v_m}\right)^2\right]
\end{eqnarray}

\noindent $v_m$ is the mean velocity, $S_{\|}$ is the parallel speed ratio and $P(v)$ is normalized. A $v^3$ pre-factor, usually included \cite{HaberlandRSI85}, has been omitted for two reasons: firstly, it has small effects if $S_{\|}$ is large, and secondly, the velocity distribution is modified by the transmission of the interferometer, because Bragg diffraction is velocity selective. In our experiment, with argon as the carrier gas \cite{MiffreEPJD05,MiffrePRA04,MiffreJCP05}, the typical values of $v_m$ and $S_{\|}$ are  $v_m\approx1062$ m/s and $S_{\|} \approx 7 $, corresponding to a distribution full width at half maximum $\Delta v_{FWHM}^{sup}=2\sqrt{\ln 2}v_m/S_{\|} \approx 250$ m/s.

For the laser-atom interaction, it is necessary to minimize Doppler broadening, so as to selectively excite the chosen hyperfine components and not the other ones, and to pump all the velocity classes with almost the same efficiency: this is possible if the laser beams are perpendicular to the atomic beam. The laser-atom interaction time is then given by $t_{int} \approx 2 w_0/v_m \approx 10$ $\mu$s, where  $w_0$ is the waist of the laser beams, $w_0 \approx 5$ mm in our experiment.

The saturation parameter $s$ is given by $s=I(\textbf{r})/I_s$ with a saturation intensity $I_s=2.56$ mW/cm$^2$ \cite{MetcalfStraten}. The number of absorption-emission cycles is $\Gamma t_{int} s/(2(1+s)) \approx 200 s/(1+s)$  and, with $s\approx 2$, the number of cycles is quite large, about $130$ and the limit $t_{int} \rightarrow \infty$ is practically reached.

The atomic beam enters the atom interferometer after a very high collimation in the $x$-direction done by a two-slit system.
The geometry of this system is such that the  width of the distribution of $v_x$ after collimation is comparable to the atom recoil velocity $v_r = \hbar k_L/M_{Li} \approx 0.09$ m/s, where $k_L = 2\pi/\lambda_L$ is the laser wave vector near the resonance line at $\lambda_L \approx 671$ nm. Because of this high collimation, optical pumping must be done before collimation, otherwise the exchanged photon momenta would completely spoil this collimation. Moreover, only a very narrow class of $v_x$ need be pumped. Due to the natural line width, a monochromatic laser interacts with a velocity class of width $\Delta v_x \approx \Gamma \lambda_L \approx 4$ m/s, considerably larger than needed, the more so that this velocity class broadens when the number of absorption-emission cycles increases.

\section{Numerical simulation of the optical pumping process}
\label{NumSimu}

The goal of this simulation is to verify that the qualitative arguments developed above are correct and also to estimate the limitations due to an imperfect circular polarization of the laser beams.

\subsection{The rate equations}

The general calculation of the optical pumping process would be quite complicated because we should consider at least $13$ Zeeman-hyperfine sublevels (8 sublevels of the $^2$S$_{1/2}$ state and 5 sublevels of the $F=2$ level of the $^2$P$_{1/2}$ state). Therefore the density matrix would have a dimension equal to $13$, with $13$ populations and $12\times 13/2= 78$ complex coherence terms. We can consider only the populations for the following reasons:

\begin{itemize}

\item the hyperfine coherence terms are negligibly small because these terms are not well excited by our pumping scheme and they are destroyed by precession at the hyperfine frequencies;

\item the Zeeman coherence terms vanish exactly when the excitation beams have a pure $\sigma^{+}$ polarization and they are weak in the presence of a weak admixture of $\sigma^{-}$ polarization. These terms are destroyed by precession at the Zeeman frequencies;

\item if the saturation parameter is small, $s \ll 1$, all optical coherence terms are small. In our experiment, the saturation parameter is close to $2$ at the center of the laser beams. However, as shown below, in the ideal case (pure $\sigma^{+}$ polarization), optical pumping is finished well before reaching the center of the laser beam and the atoms are in a dark state. In the real case, with a weak admixture of $\sigma^{-}$ polarization laser, the final state prepared by optical pumping is mostly sensitive to what occurs when the atom exits of the laser beams, i.e. in a region where the saturation parameter is small.

\end{itemize}

With these approximations, we have to consider the populations of the $13$ levels involved which are related by the following  rate equations:

\begin{eqnarray}
\dot{\sigma}_{ii} & = & -\sum_j P_{ij} \sigma_{jj}+\sum_j A_{ji}\sigma_{jj}\\
\dot{\sigma}_{jj} & = & \sum _i P_{ij} \sigma_{ii}-\sum A_{ji} \sigma_{jj}
\end{eqnarray}

\noindent where $i$ is a short-hand for $F,m_F$ and $j$ for $F', m'_F$. $P_{ij}$ describes the laser induced transition probability per unit time between sublevels $i$ and $j$ and $A_{ji}$ is the Einstein coefficient for spontaneous emission from $j$ to $i$. $A_{ji}$ and $P_{ij}$ can be expressed as a function of the transition dipole moment:

\begin{equation}
A_{ji} \propto  \left| \langle F',m_F'|d^1_{q}|F,m_F \rangle \right|^2
\end{equation}

\noindent where $d^1_{q}$  is the irreducible tensorial components of the dipole operator,$q = m_F'-m_F$. Obviously, $\sum_i A_{ji} = \Gamma$. $P_{ij}$ is a function of the laser polarization vector $\mathbf{e}$ and the local laser power density $I(\mathbf{r})$. By a simple generalization of the equations established for a two level system \cite{cohentannoudji88}, one gets:

\begin{equation}
P_{ij} =\frac{\pi I(\textbf{r})}{2 \hbar^2}|\langle ^2S_{1/2},F,m_F|\ \textbf{d}\cdot\textbf{e}\ |^2P_{1/2},F',m_F'\rangle|^2
\end{equation}

\noindent where $I(r)$ is the local laser intensity. In this expression, we have neglected Zeeman splittings and Doppler effect: Zeeman splittings because we assume that the magnetic field is small, with Zeeman splittings smaller than the natural linewidth $\Gamma$ and Doppler effect because we want to pump a very narrow velocity class with $\left| v_x\right| < v_r$. Finally, we have neglected power broadening for the same reasons we have neglected optical coherence terms. We use the Wigner-Eckart theorem to get the transition dipole matrix elements:
\begin{equation}
\begin{array}{l}
|\langle S_{1/2},F,m_F|d^1_{q}|²P_{1/2},F',m_F'\rangle|^2=
  \frac{4}{6}\left(2F+1\right)\left(2F'+1\right)  \\[0.5cm]
\begin{pmatrix}
F &1&F'\\
-m_F&q&m_F'\\
\end{pmatrix}^2
\left\{ \begin{array}{ccc}
F&1&F'\\ 1/2&3/2&1/2\\
\end{array} \right\}^2
|\langle ^2S\parallel\mathbf{d}\parallel\ ^2P\rangle|^2
\end{array}
\end{equation}
\noindent We thus get 13 linear differential equations and we use the Runge-Kutta-Fehlberg method \cite{numerical-recipes} to integrate this system and to perform numerical simulations.

\subsection{Pumping in the ideal case}

\begin{figure} [h!]
\includegraphics[width=8cm]{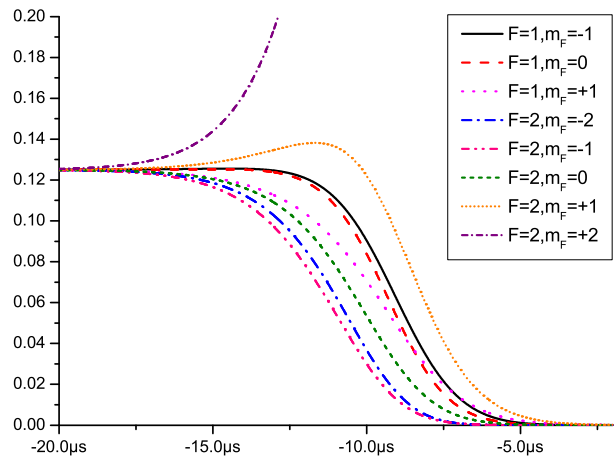}
\caption{(Color on line) Variations of the populations of the ground state sublevels during the interaction with the laser beams.
The time origin is taken when the atom is at the center of the laser beams in $z=0$ and with the velocity $v=1000$ m/s, the local laser power density is given by $ I(t)\propto \exp\left(-2(t/5)^2\right)$ with $t$ in microseconds. Pumping is almost completed  before the atoms have reached the laser beam centers at $t=0$ $\mu$s. \label{fig2}}
\end{figure}

We first consider a lithium atom propagating along the  $z$-axis with a velocity $v=1000$ m/s. Both laser beams are perfectly circularly polarized and they are described by Gaussian beams propagating along the $x$-axis. Their local intensity $I(\mathbf{r})$ is given by:
\begin{equation}
I(\mathbf{r}) = I(x=0, y=0, z)= I_0 \exp\left[-\frac{2z^2}{w_0^2}\right]
\end{equation}
\noindent where $w_0$ is the waist radius taken equal to $w_0= 5$ mm. The intensity $I_0$ is related to the laser beam power $P$ by $I_0 = P/(\pi w_0^2)$ and, in the calculations, we use the same power $P= 5 $ mW for the two laser beams. Initially, the atomic population is equally distributed among the 8 ground state sublevels. Figure \ref{fig2} plot the populations of the sublevels of the ground state as a function of time $t$. In this ideal case, the optical pumping result is essentially complete and the pumping efficiency remains excellent if the atom velocity is increased. The populations of the excited state sublevels remains always small, being at most $1$\% of the total atomic population, because most of the optical pumping is completed well before the atoms have reached the laser beam centers.

\section{Main limitations of the optical pumping efficiency}

The optical pumping scheme used here is expected to be very efficient because the population accumulates in the $F=2,m_F=+2$ sublevel which is a dark state. We discuss now the main limitations of the pumping efficiency in this case.

\subsection{Limitations due to experimental defects}

\begin{figure} [h!]
\includegraphics[width=6cm]{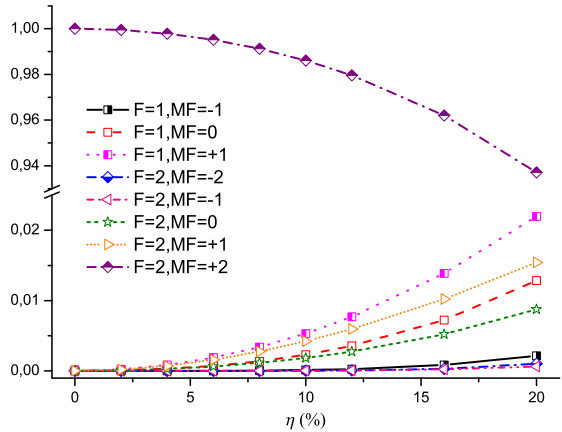}
\caption{(Color on line) Final atomic populations of the $^2$S$_{1/2}$ sublevels as a function of the fraction $x$ of laser intensity in the $\sigma^{-}$ polarization. \label{fig3}}
\end{figure}

In the actual experiment, the polarizations of the laser beams are imperfect: the polarizer or the quarter-wave plate may have small defects; the quarter-wave plate axis is not perfectly adjusted; the angle between the laser beam axis and the magnetic field is not exactly zero; the windows crossed by the laser beams when entering the vacuum chamber are slightly birefringent. We have analyzed the magnitudes of these effects and the window birefringence is likely to be dominant. If we tried to adjust the polarization to get a $\sigma^{+}$ transition, these defects introduce a weak intensity in the $\sigma^{-}$ and possible $\pi$ transitions. Nevertheless, the laser beams propagation axis and the magnetic field make an angle smaller than $20$ mrad, the fraction of the intensity in $\pi$ polarization is probably considerably weaker than in $\sigma^{-}$ polarization. Therefore, we tested the sensitivity of the optical pumping efficiency with a fraction $\eta$ of the laser intensity in $\sigma^{-}$ polarization and a fraction $(1-\eta)$ in $\sigma^{+}$ polarization. The results of the calculation are plotted as a function of this fraction in fig. \ref{fig3}.

The result is that the population not transferred in the $F=2,m_F=+2$ sublevel is small and it increases quadratically with $\eta$ for the sublevels which are repopulated after absorption of one photon from $F=2,m_F=+2$ sublevel and like $\eta^3$ for the other ones. We performed a similar calculation assuming laser beams with a top-hat profile and, for the same fraction $\eta$, the population not transferred in the $F=2,m_F=+2$ sublevel is considerably larger. We explain qualitatively this result as follows: when an atom goes out of a Gaussian beam, the last absorbed photons are of the $\sigma^{+}$ polarization with a probability equal to $(1-\eta)$ and the atom exits of the beam almost as if the beam was purely $\sigma^{+}$. In a top-hat beam, the atom internal distribution reaches a steady state which is conserved when the atom goes out abruptly of the laser beams.

\subsection{Limitations due to atom density effects}
\label{someresultsPO}

Atom-atom collisions and radiation trapping are the main effects which are dependent on the atom density and which limit the optical pumping efficiency.

In our experiment, the excited state population keeps very weak and the dominant collisions should be spin-exchange collisions between ground-state atoms \cite{HapperRMP72} but they should play a negligible role. Spin-exchange collisions play no role when the gas is fully pumped in a single sublevel with complete spin polarization, which is the case in our experiment. Moreover, collisions in a supersonic beam after the skimmer are very rare with a mean number of collisions per atom substantially less than $1$ and the dominant collisions involve the carrier gas, which is a noble gas in its $^1$S$_0$ ground state, with no possibility of spin exchange. Lithium-lithium collisions are considerably less numerous, by a factor of the order of the lithium to carrier gas density ratio in the oven, about $3\times 10^{-3}$. Finally, as spin-exchange collision cross-section and total atom-atom cross-section have comparable magnitudes, an important fraction of collisions leading to spin exchange also induce a reorientation of the relative velocity vector and such a collision ejects the atom out of the atomic beam with a large probability.

Radiation trapping also reduces the pumping efficiency \cite{HapperRMP72}: when an atom absorbs a fluorescence photon emitted by another atom, this photon is largely depolarized, so that this absorption induces a leak in the pumping process. D. Peterson and L.W. Anderson \cite{PetersonPRA91} have studied the effect of radiation trapping on the polarization of an optically pumped alkali atomic beam. Their calculation, which is simplified, does not apply exactly to our case, as they consider an effusive sodium beam with a broad longitudinal velocity distribution and a negligible transverse velocity distribution while our supersonic beam has a narrower longitudinal velocity distribution but their pumping scheme is similar to ours, with the atoms transferred in a dark state. They find that the important parameter is the product of the atom density $n$  by the beam diameter $D$ in the pumping volume. In the case of a transverse optical pumping,  optical pumping efficiency starts to decrease when $n D > 10^{11}$ atoms/cm$^2$ and we have calculated that the narrower velocity distribution of our beam should decrease this threshold by a factor $3$. Using the results of A. Miffre \cite{MiffrePhD}, we estimate the atom density in the optical pumping region:

\begin{equation}
n_{Li} =\frac{I_{Li}}{v_m z^2} \approx 5 \times 10^{9} \mbox{  atoms/cm}^3
\end{equation}

\noindent with the calculated beam intensity $I_{Li} \approx 4 \times 10^{16}$ atoms/(s.sr), the measured mean velocity
$v_m=1062$ m/s and the distance $z$ from the nozzle to the center of the pumping volume, $z \approx 9$ cm. At this place,
the atomic beam diameter $D$ is about $D \approx 0.8$ cm, so that $n_{Li}D \approx 4 \times 10^{9}$ atoms/cm$^2$ well below the lowered threshold near $n D > 3\times 10^{10}$ atoms/cm$^2$: as a consequence, we expect radiation trapping to have negligible effects in our experiment.

We may also use some previous experimental results to test the calculation of D. Peterson and L.W. Anderson \cite{PetersonPRA91}.
B.P. Masterson \textit{et al.} \cite{MastersonPRA93} have pumped a very intense cesium beam with a rectangular cross section $2.5$ $\times0.5$ cm$^2$, with several tricks to reduce radiation trapping. From their data, we estimate the density in the pumping volume  $n \approx 10^{10}$ atoms/cm$^3$, so that $n D \approx 10^{10}$ atoms/cm$^2$ and they achieved less than $2\times 10^{-4}$ of the population left in the depleted $F$ level and up to $95$\% of the population in a single $F,m_F$ sublevel. G.W. Schinn \textit{et al.} \cite{SchinnJOSAB91} have pumped a sodium beam with a  $D=0.4$ cm diameter: the spin polarization, which was $97$\% with a density $n \approx 2\times 10^{10}$ atoms/cm$^3$, decreased to $\approx 94$\% with a density $n \approx 2\times 10^{11}$ atoms/cm$^3$, in agreement with reference \cite{PetersonPRA91}.

\section{Experimental setup}

\subsection{The atomic beam and the atom interferometer}
\begin{figure} [h!]
\includegraphics[width=8cm]{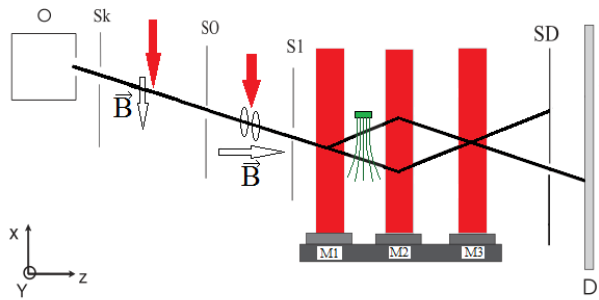}
\caption{(Color on line) Schematic top-view of our Mach-Zehnder atom interferometer. The lithium supersonic beam, after passing the skimmer (SK) is strongly collimated by two slits (S0 and S1). It is then diffracted in the Bragg regime by three laser standing waves, produced by reflecting laser beams on the mirrors M$_1$, M$_2$ and M$_3$. The interferometer thus produces two output beams with complementary fringe signals. One of the output beams is selected by a slit (SD) and detected by a Langmuir-Taylor "hot-wire" detector (D). The optical pumping volume is located $5$ cm after the skimmer (SK), with the laser beams crossing the supersonic beam at right angle (in order to visualize the diffracted beams, the Bragg angle, equal to $80$ $\mu$rad for a lithium velocity near $1000$ m/s, is exaggerated). The deflection experiment is made by illuminating the atomic beam by a laser beam, between the collimation slits, in the presence of an homogenous magnetic field parallel to the $z$-axis. The magnetic dephasing experiment is made by applying a magnetic field gradient, between the first and second diffraction standing waves.     \label{fig4}}
\end{figure}

Our experimental set-up is described in \cite{MiffreJCP05}. Here are its main features and some details concerning the optical pumping setup (see fig. \ref{fig4}). A supersonic lithium beam is produced by seeding lithium in a noble gas, the oven being heated to $800$ $^{\circ}$C to insure a sufficient density of lithium. By using different carrier gases, the mean beam velocity can be varied: $744$ m/s with krypton, $1062$ m/s with argon and $1520$ m/s with neon \cite{LepoutreEPJD11}. After passing the skimmer, the atomic beam is strongly collimated by two slits to achieve a transverse velocity distribution with a width comparable to lithium  recoil velocity ($\approx$ 9 cm/s). This atomic beam is diffracted by three quasi-resonant standing laser waves in the Bragg regime, which produces only two beams. For the present experiments, we use first order diffraction and we get a Mach-Zehnder atom interferometer with two output beams carrying complementary fringe signals: one of these two beams is selected by a slit and detected by a Langmuir-Taylor "hot-wire" detector. The interferometer arms are spatially separated and their maximum distance, which is reached when crossing the second laser standing wave, decreases when the beam velocity increases. It is equal to $143$ $\mu$m for $v_m= 744$ m/s, to $100$ $\mu$m for $v_m= 1062$ m/s and to $70$ $\mu$m for $v_m=1520$ m/s.

The standing wave laser is detuned ($\Delta \nu \approx $2 GHz) on the blue side of the $^2S_{1/2} \rightarrow ^2P_{3/2}$ transition of $^7$Li so that the diffraction probability of $^6$Li is very small and, in addition, its natural abundance is small ($7.5$\%), so the contribution of $^6$Li to the interference signal is fully negligible. The interferometer signal $I$ is given by:

\begin{equation}
 I=I_0 \left(1 + \mathcal{V} \cos{(\varphi_d + \varphi_p)}  \right)
\end{equation}
\noindent where $I_0$ is the mean intensity and $\mathcal{V}$ the fringe visibility. $\varphi_p$ is the phase induced by perturbations applied on the interferometer arms and $\varphi_d$ is the phase due to the diffraction process, $\varphi_d= 2 k_L (x_1-2x_2+x_3)$, where $k_L$ is the laser wave vector and $x_i$-position of the standing wave mirror M$_i$. The fringe signal is scanned by varying the position $x_3$ with a piezo-actuator on the mirror $M_3$. We can control the position $x_3$ with an optical Michelson interferometer. When using argon as the carrier gas, the typical values of $I_0$ and of $\mathcal{V}$ are $I_0 \approx 6 \times 10^4 $ atoms/s and $\mathcal{V}\approx 70$\%, .

\subsection{The magnetic field in the pumping, deflection and dephasing regions}
The whole interferometer experience the Earth magnetic field: the field is modified by the presence of large piece of steel used to support the vacuum chambers but its magnitude is a few $10^{-5}$ T everywhere. In addition to this field, we produce magnetic fields in three different regions.

In the optical pumping region, just after the skimmer and before collimation of the atomic beam, we produce an homogeneous magnetic field by three pairs of square Helmholtz coils \cite{MerrittRSI83} with $80$ windings each. The square sides are equal to $50$, $40$ and $30$ mm, so that each smaller one fits into a bigger one. The measured laboratory field components are equal to $B_x= -3.0 \times 10^{-5}$ T, $B_y = -2.3\times 10^{-5}$ T  and $\left|B_z\right| < 10^{-5}$ T. The Helmholtz coils are used to cancel $B_y$ and to produce along the $x$-axis an additional contribution $B_x = \pm 3.8 \times 10^{-4}$ T. This means that when we reverse the current we do not exactly reverse $B_x$ which is equal either to $B_x = - 4.1 \times 10^{-4}$ T or to $B_x = + 3.5 \times 10^{-4}$ T.

The laser deflection experiment is made by shining a laser beam perpendicular to the atomic beam, between the two collimation slits and an homogenous magnetic field is needed in order to separate the various Zeeman components so that this deflection process can be state-selective. The field is chosen parallel to the $z$-axis and it must be strong and homogenous. Two $40$ mm diameter Helmholtz coils, with $38$ windings each, produce a field $B_z/I= 18.8 \times 10^{-4}$ T/A where $I$ is the current in the coil.

The Zeeman phase shift experiment uses magnetic coils, originally designed to compensate the gradient of the magnetic field used for the measurement of the He-McKellar-Wilkens phase \cite{LepoutrePRL12}. This $30$ mm diameter coil is located roughly at mid-distance between the first and second laser standing waves; it has $9$ windings and, with a current $I_C$ (A),  it produces a field $B_x \approx (3 \times 10^{-4} I_c) $ T on the interferometer arms, with a gradient of the order of $\partial B_x/\partial x  \approx (3 \times 10^{-3} I_c)$ T/m.

\subsection{Laser system for optical pumping}

\begin{figure} [h!]
\includegraphics[width=8cm]{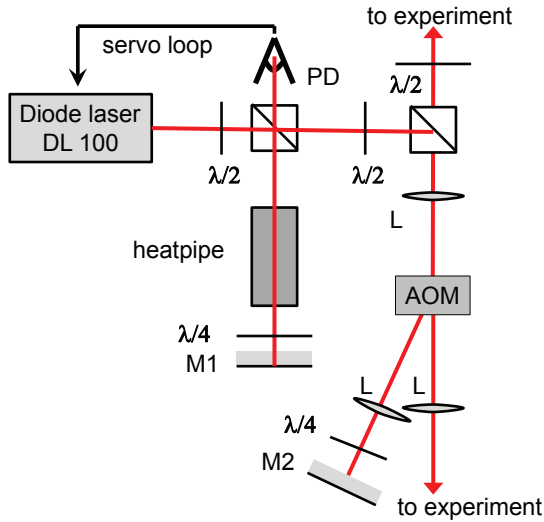}
\caption{(Color on line) Production of the laser beams needed for optical pumping. The beam emitted by a TOPTICA DL 100 diode laser is split in two beams: a weak beam (below $2$ mW) is used to generate Doppler-free signals by saturated absorption spectroscopy on the $^2$S$_{1/2}$,$F=1 \rightarrow ^{2}$P$_{1/2}$, $F=2$ transition in a heat-pipe oven. The laser frequency is locked on these signals. The main beam is sent through an acousto-optic modulator (AOM) operating a a frequency $\nu_{RF}$ near $401.5$ MHz. The direct zeroth-order beam is sent to the experiment. The first-order beam is sent back through the AOM to produce a second beam, red-shifted by $2\nu_{RF}$ which is also sent to the experiment.
\label{setupPO}}
\end{figure}

We use an extended cavity single-frequency diode laser from TOPTICA (model DL 100) to produce the laser beams used to pump $^7$Li. The laser set-up is shown on fig. \ref{setupPO}. We use frequency modulation and saturated absorption spectroscopy in a heat pipe oven to lock the laser on the $^2$S$_{1/2}$,$F=1 \rightarrow ^{2}$P$_{1/2}$, $F'=2$ transition. The laser beam is sent to an acousto-optic modulator (AOM) operating with $\nu_{RF}\approx 400$ MHz. By tuning the RF power sent to the AOM, we get two laser beams of roughly equal powers. One of these beams is the zeroth-order beam which is not shifted in frequency and is resonant with the $F=1 \rightarrow F'=2$ transition. The other beam is the first-order beam which is diffracted twice and so frequency shifted by $2\nu_{RF} \approx 803.5$ MHz, this beam is resonant with the $F=2 \rightarrow F'=2$ transition.
Both beams are sent to the optical pumping region and as they must have the same circular polarization, it is difficult to combine them in a single beam without large power losses: we have chosen to propagate the two beams in the $x-y$ vertical plane, both perpendicular to the atomic beam and making small angles near $\pm 0.5$ mrad with the $x$-axis. The two laser beams go through the same polarizer and the same zeroth-order quarter-wave plate before entering the vacuum chamber and they converge in the pumping volume where their beam waists are around $5$ mm and their power is about $5$ mW per beam. We optimize $\nu_{RF}$ by maximizing the laser power absorbed by the atomic beam.

\section{Experimental results and analysis}
\label{ExpResults}

\subsection{Possible tests of the internal state distribution}
\label{p6}
Stern-Gerlach magnets or hexapole magnets have been used to measure the electronic spin polarization \cite{BaumZPD91}. This technique does not give access to the population of individual $F,m_F$ sublevel. The usual way of measuring the complete distribution over the  $F,m_F$ sublevels is based on laser induced fluorescence in the presence of a magnetic field large enough to separate the Zeeman components of the transition:  G.W. Schinn \textit{et al.} \cite{SchinnJOSAB91} have used  $B \approx 2\times 10^{-2}$ T. Another technique is to induce radio-frequency or microwave transitions between the hyperfine levels in the presence of a magnetic field sufficient to resolve the Zeeman components of the transition. The radio-frequency or microwave transitions are then detected by measuring the population of the populated level by laser induced fluorescence. As the transitions between hyperfine levels are very narrow, a weaker magnetic field can be used, for instance a field of $B \approx 7\times 10^{-4}$ T was used by B.P. Masterson \textit{et al.} \cite{MastersonPRA93}.

In our experiment, only a narrow velocity class near $v_x =0$ is optically pumped which complicates the measurement of the population distribution over the sublevels. If we measure the atomic populations before collimation, the signal will be sensitive to velocity classes which may not be completely pumped and which are eliminated by the collimation process. If we make the measurement on the atomic beam after collimation, the atomic flux is small, below  $10^6$ atoms/s, and the beam velocity is of the order of $1000$ m/s: then, the expected fluorescence signal is very low and laser stray light will be a problem. Because of these considerations, we did not tried to use any of these two techniques. Instead, we have used two measurements techniques for which the signal is measured on the Langmuir-Taylor "hot-wire" detector:
\begin{itemize}
\item the first technique is laser deflection of the beam, in the presence of a magnetic field sufficient to resolve the Zeeman components of the transition. Atom deflection has already been used by Gould \textit{et al.} \cite{GouldPRA87} to characterize a sodium beam.

\item the second technique is based on the modification of the atom interferometer signals by a weak magnetic field gradient \cite{SchmiedmayerJPII94,GiltnerPhD96,JacqueyEPL07,LepoutrePhD11}. The idea is that the magnetic field gradient induces a phase shift which varies with the $F,m_F$ sublevel so that the phase and visibility of the interference fringes are sensitive to the population distribution over these sublevels.
\end{itemize}

\subsection{Laser deflection of the atomic beam}
We assume that the magnetic field is strong enough to lift the Zeeman degeneracy, so that the process is selective in $F$ and $m_F$. The collimated atomic beam has a very narrow distribution of $v_x$ and, if a resonant laser crosses at right angle the atomic beam, an atom will absorb a laser photon and emit a spontaneous photon, thus going to one of its ground state sublevels. The momentum of the spontaneous photon is random so that the total modification of $v_x$ is between $0$ and $2 v_r$, with a mean value equal to the recoil velocity $v_r \approx 9$ cm/s. This is sufficient to reduce substantially the probability that the atom reaches the detector. Moreover, after spontaneous emission, the atom has a certain probability of being back in the same $F,m_F$ state and to be re-excited a second time, which induces a second  momentum transfer identical to the first one. A complete model of this multi-step process is needed if one wants to predict the efficiency of laser deflection.

\begin{figure} [h!]
\includegraphics[width=8cm]{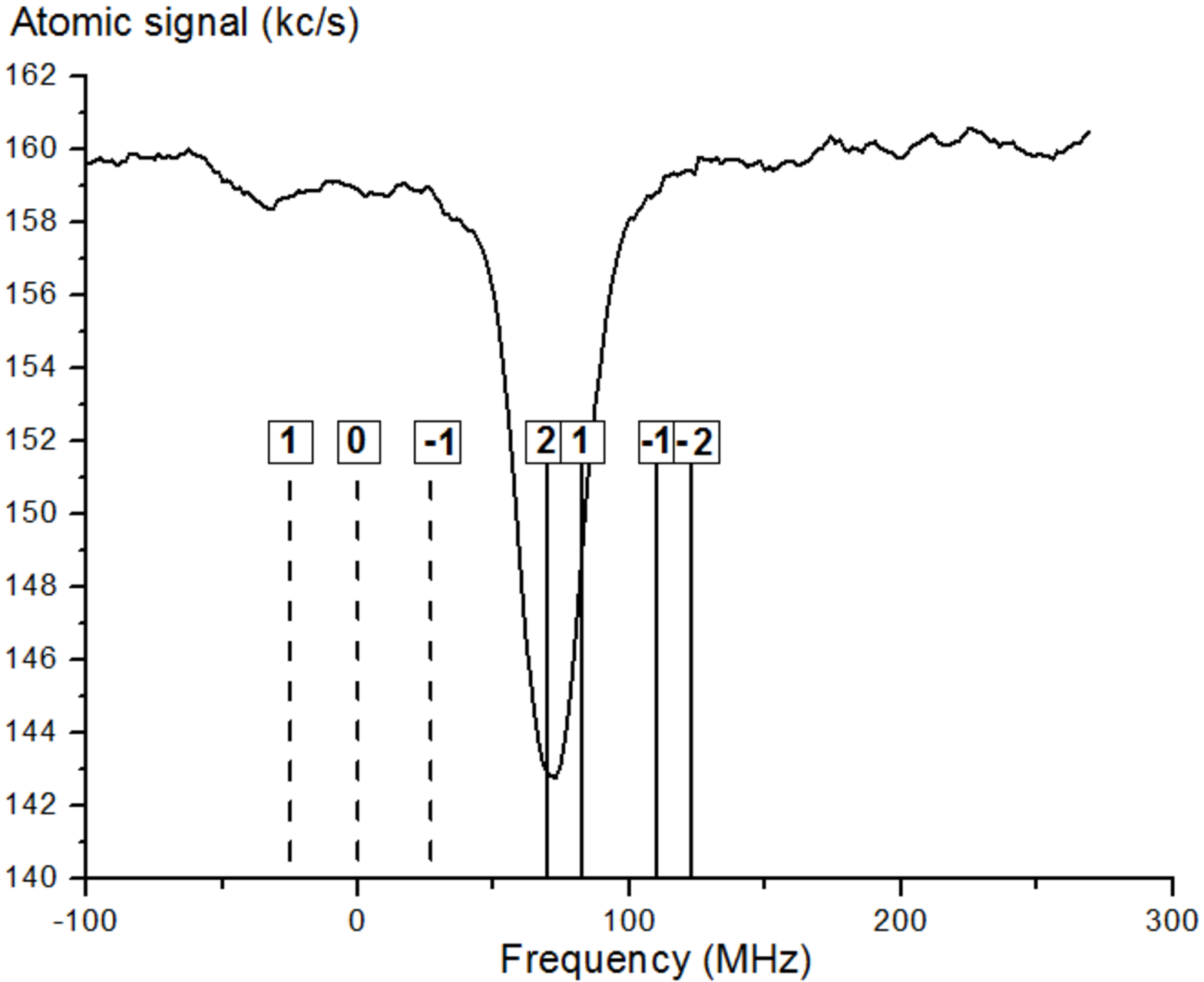}
\includegraphics[width=8cm]{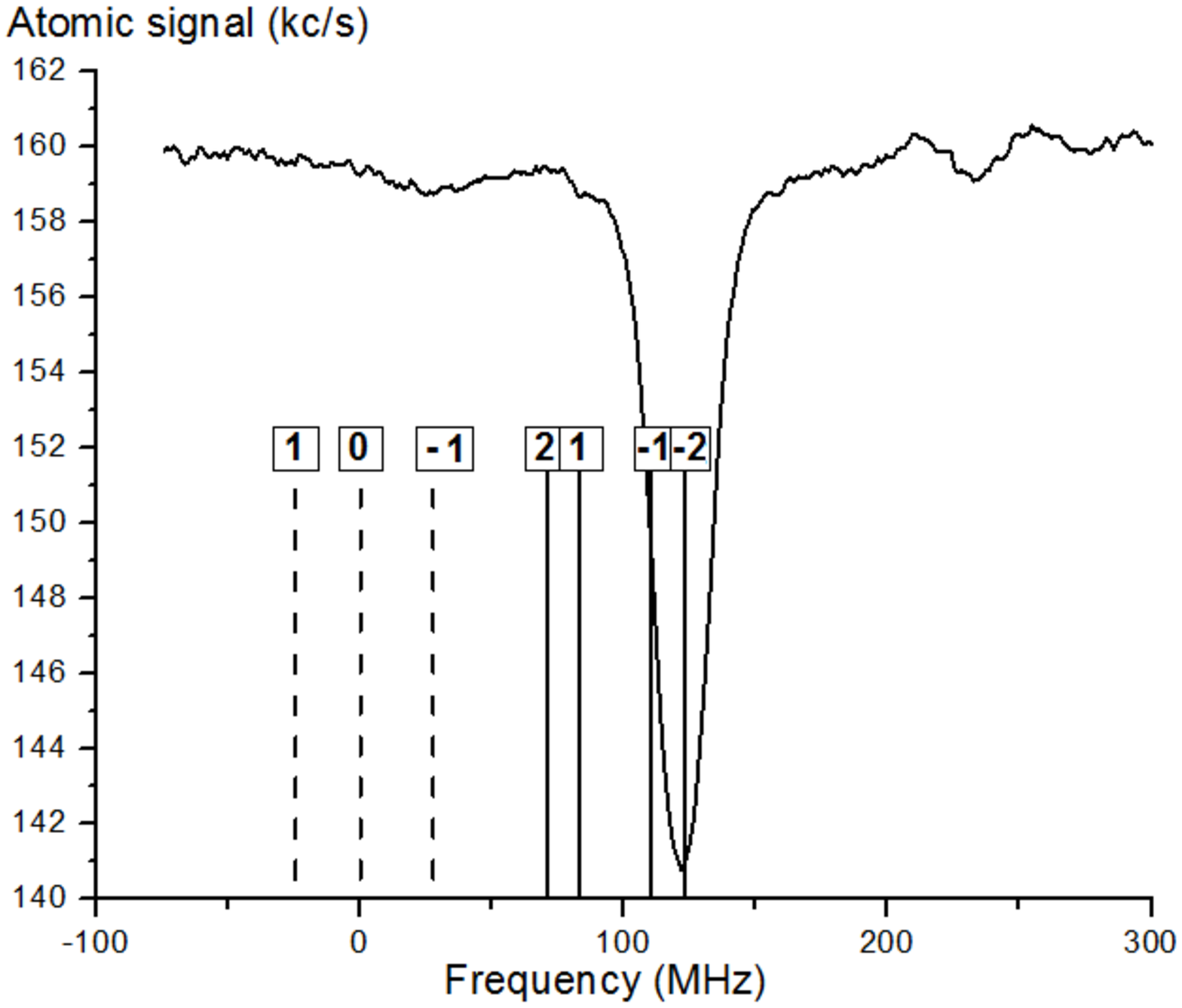}
\caption{Laser deflection experiment: the measured atomic beam intensity is plotted as a function of the laser frequency in the range of the $F=2 \rightarrow F'=1,2$ transitions in the presence of a magnetic field $B= 28.4\times 10^{-4}$ T (dashed lines for $F'=1$, solid lines for $F'=2$). We have not used a larger field because the two transitions start overlapping and this situation extends to the maximum available field. Upper panel: pumping in the $m_F=+2$ sublevel; lower panel: pumping in the $m_F=-2$ sublevel. The beam intensity is counted every $0.1$s and we have reduce the noise by a sliding average over $14$ data points. This experiment was made with a beam mean velocity $v_m =1062$ m/s (argon as a carrier gas). The intensity dip is slightly larger for the $m_F=-2$ sublevel than for the $m_F=+2$ sublevel, suggesting that the pumping efficiency is slightly better for $m_F=-2$. \label{spectro}}
\end{figure}

We choose the D1 line $^2$S$_{1/2}\rightarrow ^2$P$_{1/2}$ for this experiment, for the same reasons we have chosen it for optical pumping. We excite the $F=1,2 \rightarrow  F'=1,2$ transitions with a laser propagating along the $x$-axis. The laser polarization is linear, parallel to the magnetic field $\mathbf{B}$ which is along the $z$-axis. With $\pi$ polarization
the selection rule is $\Delta m_F= m'_F-m_F=0$ and we should observe only $12$ lines (the transitions involving $m_F=0$ and the same $F$-values are forbidden). We record the hot-wire detector signal as a function of the laser frequency and we expect to see a dip in the signal each time the laser is resonant with an excitation transition, provided that the lower level population is not vanishing. Ideally, we should compare the signals recorded with an optically pumped beam to those recorded without optical pumping. Unfortunately, in the non pumping case, all the hyperfine sublevels are equally populated and the signal to noise ratio on each individual line is small. In the presence of optical pumping shown in fig. \ref{spectro}, only one line is well detected, corresponding to the populated level $F=2, m_F= \pm 2$ (or $-2$) and the absence of the lines corresponding to the other levels proves that their population is considerably weaker. The dip in the signal is rather small because it was necessary to reduce the laser power density near $1$ mW/cm$^2$ to prevent an excessive broadening of the line. From the observed signals, for all the levels for which the detection line is well resolved from the line due to the populated sublevel (for all the sublevels except the  $F=2, m_F=+1$ sublevel (or $-1$)), one can deduce that the population cannot be larger than about $3$\% of the total population but it is impossible to make a precise measurement of the atomic distribution over the hyperfine sublevels.

\subsection{Magnetic dephasing of the atom interferometer signals: theory}
An inhomogeneous magnetic field on a matter wave interferometer modifies the phase of the interference fringes. Such an experiment \cite{aharonovo67,bernstein67} was proposed to test the sign reversal of a spin $1/2$ wavefunction by a $2\pi$ rotation. It was performed in 1975 by H. Rauch and co-workers \cite{rauch75} with a perfect crystal neutron interferometer (for a review \cite{rauch00}). Similar experiments can be done with atom interferometers: the application of a magnetic field gradient induces a phase-shift $\varphi_{Z, F, m_F}$ which is a function of $F, m_F$ and of the atom velocity. The fringe signal is given by:

\begin{eqnarray}
\label{a10}
\frac{I}{I_0} &=&  \sum_{F,m_F} P(F,m_F)\nonumber \\ && \left<1 + \mathcal{V}_0\cos\left(\varphi_d + \varphi_{Z,F,m_F} + \varphi_{Sagnac}\right)\right>
\end{eqnarray}
\noindent where $\left<...\right>$ refers to the average over the velocity distribution $P(v)$ given by eq. (\ref{a2}). We have also taken into account the Sagnac phase shift due Earth rotation \cite{JacqueyPRA08}: $\varphi_{Sagnac}$ is proportional to $1/v$ and it is equal to $0.688$ rad for $v= 1000$ m/s. As the Zeeman phase shifts $\varphi_{Z,F,m_F}$ are very different for the various sublevels, the corresponding contributions to the fringe signal are no more in phase and the fringe visibility exhibits a series of minima and revivals when the gradient increases \cite{SchmiedmayerJPII94,schmiedmayer97,giltnerPHD96,JacqueyEPL07}.

The phase-shift  $\varphi_{Z,F,m_F}$ is easy to calculate if, as in previous studies, we assume an adiabatic
behavior. The direction of the magnetic field $\mathbf{B}$ is slowly varying in space so that the projection $m_F$ of the $\mathbf{F}$ angular momentum on an axis parallel to the local field is constant. The interferometer arms are very close to the $z$-axis and we note $\delta x \left(z \right)$ the distance between the two arms, the Zeeman phase shift is related to the Zeeman energy shift $U_{Z,F,m_F}$ of the $F,m_F$ sublevel by:

\begin{equation}
\varphi_{Z,F,m_F} = -  \int \frac{\partial U_{Z,F,m_F}}{\partial B} \frac{\partial B}{\partial x} \frac{\delta x \left(z \right)dz}{\hbar v}\label{PhiZ}
\end{equation}

\noindent $\varphi_{Z,F, m_F}$ varies with the atom velocity like $1/v^2$, with one $v$ factor obvious in eq. (\ref{PhiZ}) and the other one hidden in the distance between the two interferometer arms $\delta x \left(z \right)$. If the field is low enough, Zeeman effect is linear and we can write:

\begin{eqnarray}
\label{z7}
\varphi_{Z,F,m_F} & =&  g_Fm_F J_1= g_F m_F A_{J_1} I_C \nonumber \\
 J_1 &=&A_{J_1} I_C  =  \frac{\mu_B}{\hbar v} \int \frac{\partial B}{\partial x} \delta x(z)  dz
\end{eqnarray}

\noindent where $I_C$ is the current in the coil producing the field gradient. The ground state Land\'e factors $g_F$ are equal to $g_{1} \approx -1/2$ and $ g_{2} \approx 1/2$ and, throughout this discussion, we will neglect the very small nuclear contribution to Zeeman effect. However, the eq. \ref{z7} is too simple as it neglects the laboratory field which dominates the field due to the magnetic coil when the current $I_C$ is small. Moreover, the gradient along the $x$-axis of the laboratory field is not negligible. It is easy to show that, for moderate and strong fields, the effect of the laboratory can be approximately represented by an offset $I_{0,C}$ of the coil current $I_{C}$ and a supplementary phase shift $J_{0,C}$ due to the gradient outside the region where the magnetic field created by the magnetic field dominates the laboratory field. Taking these corrections into account, eq. \ref{z7} becomes:

\begin{eqnarray}
\label{phifitlinear}
\varphi_{Z,F,m_F} & =&  g_F m_F \left( A_{J_1}  \left|I_{C}-I_{0,C} \right| + J_{0,C}\right)  \nonumber \\
\mbox{with  } J_{0,C} &=& \frac{\mu_B}{\hbar v} \int \frac{\partial B_0}{\partial x} \delta x(z)  dz
\end{eqnarray}

\noindent The magnetic field produced by the magnetic coil is strong enough to partially uncouple the ground state hyperfine structure and nonlinear Zeeman effect appears. We take this effect into account with the second order term in $B$ in the expansion of $\varphi_{Z,F,m_F}$. For this term which is rather small, we do not make any correction for the laboratory field:

\begin{eqnarray}
\label{z8}
\varphi_{Z,F,m_F} & =&  g_F m_F \left( J_1 \left|I_{C}-I_{0,C} \right| + J_{0,C}\right) + \nonumber \\
 & & \pm  (1- \frac{m_F^2}{4} ) A_{J_2}  I_C^2   \nonumber \\
A_{J_2} I_C^2 &=& \frac{\mu_B^2}{\hbar v \Delta E} \int  B \frac{\partial B}{\partial x} \delta x(z)  dz
\end{eqnarray}

\noindent where $\Delta E$ is the ground state hyperfine splitting and the $\pm$ sign is associated to $F= I \pm 1/2$. We assume that a Zeeman phase shift of the third order $J_3$ is fully negligible since the coil is too small and the applied current are too weak.

\subsection{Magnetic dephasing of the atom interferometer signals: the signals and their analysis}

\begin{figure}
\includegraphics[width=8cm]{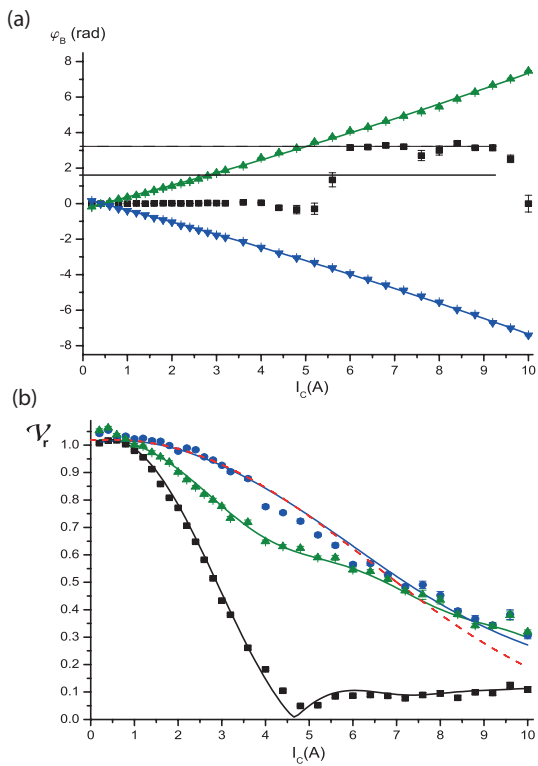}
\caption{(Color on line) (a) Phase shift $\Delta \phi = \phi(Ic)-\phi(0)$ versus the current $I_c$. The experimental data points are represented by (black) squares in the absence of optical pumping. In the presence of optical pumping, the data points are (green) triangles when pumping in the  $F=2, m_F=+2$ sublevel and (blue) round dots when pumping in the $F=2, m_F=-2$ sublevel. (b) The relative fringe visibility $\mathcal{V}_r(I_C)$ is plotted as a function of the current $I_C$ in the compensator coil. The dashed line is a simulation for $100$\% optical pumping into the $m_F=+ 2$ state and the full lines are the best fits to the data (see text). Around $I_c = 5$ we observe the velocity revival and and a phase of $\pi$ rad for the non-pumping case.\label{VisiNe}}
\end{figure}

The atom interferometer signals versus the applied magnetic field strongly depend whether the atoms are optically pumped in a well-defined $F$, $m_F$ sublevel or not. Therefore, we can estimate the efficiency of the optical pumping using a fit of the data.

In the fig. \ref{VisiNe}, both the phase shift $\Delta \phi = \phi(Ic)-\phi(0)$ and the relative fringe visibility $\mathcal{V}_r(I_C)= \mathcal{V}(I_C)/ \mathcal{V}(I_C=0)$ are plotted versus the current coils. We show three experimental cases: when the optical pumping is turned off (atoms are supposed to evenly populate the $F$, $m_F$ sublevels) and when optical pumping is adjusted to prepare the atoms either in $F=2$, $m_F=2$ or $m_F=-2$.
In the absence of optical pumping, the phase shifts induced by atoms in different sublevels cancel out and the measured phase shift is equal to 0. With the optical pumping in $F=2$ $m_F = +2$ (or $m_F = -2$), the measured phase shift scales linearly with the current $I_c$. In the experiments $|Ic|< 10A$, so the effect of a residual population in the unwanted sublevels is very weak and does not impact the phase shift measurement. In consequence, the phase shift signal is not suitable for the optical pumping evaluation.

On the other hand, fitting the relative visibility allows for a better estimation of the population in the sublevels. If we consider a boundary case, with no optical pumping, the relative visibility exhibits a cancelation and a weak revival, due to beat notes between discrete values of the phase shifts $\varphi_Z$ according to the $m_F$ sublevels. In case of a linear Zeeman approximation, the visibility cancelation appears when $\varphi_Z = \pm m_F\times \pi$ (for details \cite{JacqueyEPL07})). For the other boundary, when we have a perfect optical pumping, a single sublevel $F=2$ $m_F=2$ (or $m_F=-2$) is populated. In this case, all the atoms with a given velocity have an identical Zeeman phase shift scaling with $1/v^2$. Therefore, the average over the velocity distribution of the atoms leads to an almost a Gaussian curve for the relative visibility. The discrepancy between the data and the Gaussian shape comes from residual population in the unwanted sublevels.

This data processing is a difficult task because we have to fit a nonlinear function with several dependent parameters. In order to estimate these parameters we need three experimental configurations: without optical pumping, with optical pumping in $F=2$, $m_F=2$ and $m_F=-2$. In the configuration with optical pumping, almost all the atoms are in $F=2,m_F=+2$ (or $F=2,m_F=-2$) so the nonlinear term in eq. \ref{z8} is too weak to have a significant impact on the fit. Therefore, the Zeeman phase shift depends on the factor $g_F \times m_F$, so we have to consider only five (actually four normalized populations) different values according to the sublevels (table~\ref{fitparametersC}).

In addition to these 4 normalized populations, we have 7 other parameters to fit the data. The first parameter is the visibility $V_0$ when the atoms does not feel a magnetic field gradient. Then we have to consider the two biggest effects: the Zeeman phase shift (equation~\ref{z8}) and the hyperfine dependence of the atom diffraction.

In order to fit the Zeeman effect, we need to know the coefficient $J_1$ (eq. \ref{z8}) which links the magnetic field experienced by the atoms along their trajectories and the current through the coils. Since this parameter change slightly with the temperature we added a scale factor. In addition, the atoms see the residual magnetic field in the vacuum chamber which is estimated by the current $I_{0C}$ for which this residual magnetic field and the field due to magnetic coils cancel. The Zeeman phase shift also depends on the velocity distribution which impacts the visibility. We can include this effect with a very good approximation by assuming a gaussian velocity distribution and using the parallel velocity distribution speed ratio $S_\|$.

The other effect is the hyperfine dependence of the laser diffraction of the atoms due to the laser detuning which is not the same for the both hyperfine levels and also because the two transitions does not have the same optical transition amplitudes. However, we can balance these two effects by setting the laser to a proper detuning. This effect is modeled in the fit process which add an additional parameter to consider an imperfect cancelation.

Finally, the fit allows for a good estimation of the populations $P(F,m_F)$ after optical pumping in $m_F=+2$ and $m_F=-2$. In the table~\ref{fitparametersC}, we report the results for the lithium beam with neon as a carrier gas. The pumping efficiency is quite good: for a $\sigma_+$ pumping ($B_x > 0$) we have $P(F=2, m_F=+2)=90.1\pm 1.2\%$ and for a $\sigma_-$ pumping ($B_x < 0$) we have $P(F=2, m_F=-2)=94.1\pm 1.7\%$.

\begin{table}[h!]
\begin{tabular}{|l|l|l|}
    \hline
    parameter      & $B_x > 0$        & $B_x < 0$   \\ \hline
    P(2,2)         & $0.000\pm 0.060$ & $0.901\pm 0.012$ \\
    P(2,-2)        & $0.941\pm 0.017$ & $0.002\pm 0.032$\\
    P(2,1)+P(1,-1) & $0.006\pm 0.011$ & $0.000\pm 0.001$ \\
    P(2,-1)+P(1,1) & $0.050\pm 0.018$ & $0.038\pm 0.011$ \\
    P(2,0)+P(1,0)  & $0.004\pm 0.028$ & $0.059\pm 0.010$ \\ \hline
   \end{tabular}
\caption{Populations $P(F,m_F)$ sublevels deduced from the fits of the visibility and phase shift data set for the lithium beam with neon as the carrier gas (mean velocity $v_m= 1520$ m/s). \label{fitparametersC}}
\end{table}

This fit supports what appears clearly in figure \ref{VisiNe}:  the measured visibility is closer to the predicted visibility for a perfect pumping when we pump in the $m_F=2$ than in the $m_F=+2$ sublevel. This result is not easy to understand as we simply change the magnetic field direction. We have considered the possibility of some depumping by Majorana transitions when the atom exits of the Helmholtz coils. In this region, $B_x$ and $B_y$ change signs. In order to change the $m_F$-value of the pumped level, we invert the current in the Helmholtz coils producing $B_x$ without changing its value, the situation is not exactly symmetric for the two pumping cases. If this explanation was correct, the depumping should be larger for the same $m_F$ level when we vary the atomic beam velocity and it should increase with this velocity. The measured populations of the $F=2,m_F=\pm 2$ level for different values of the atomic beam mean velocity are collected in table \ref{POresults} and these results do not exhibit any systematic trend with the atomic beam velocity nor with the sign of the $m_F$ value of the pumped level.

\begin{table}[h!]
\begin{tabular}{|l|c|rcr|}
    \hline
    $v_m$ (m/s)             &     $m_F$    & \multicolumn{3}{c|}{$P(2,m_F)$}          \\ \hline
    $744\pm 18$             & $+2$ & ($96$  & $\pm$ & $6$) \% \\
                            & $-2$ & ($93$  & $\pm$ & $7$) \% \\
    $1062\pm 20$            & $+2$ & ($100$ & $\pm$ & $13$)\% \\
                            & $-2$ & ($95$  & $\pm$ & $11$)\% \\
    $1520\pm 38$            & $+2$ & ($90$  & $\pm$ & $1$) \% \\
                            & $-2$ & ($94$  & $\pm$ & $2$) \% \\ \hline
\end{tabular}
\caption{Optical pumping efficiency for different atom velocities: the measured population $P(F=2, m_F)$ in the pumped sublevel is given as a function of $m_F=\pm 2$ for three different atomic beam mean velocity $v_m$.   \label{POresults}}
\end{table}

Even if we have no evidence of Majorana transitions, it would be good to improve the control on the magnetic field in this experiment:

\begin{itemize}

\item a guiding field of fixed direction and of the order of a few $10^{-5}$ T would not perturb the interferometer and it would prevent any risk of Majorana transitions. In addition, it might help to reduce the $x$-gradient of the laboratory field; this would be interesting as this gradient complicates the analysis of the Zeeman phase shifts (see eq. (\ref{phifitlinear}));

\item the magnetic coil geometry could be improved with a combination of Helmholtz and anti-Helmholtz coils of different sizes. Moreover, the cooling of this coil should be improved: presently, after a long period of operation with a large current, the thermal expansion of the coil support is sufficient to modify the field gradient at the location of the interferometer arms.

\end{itemize}

\section{Conclusion}
\label{conclusion}

This paper describes the optical pumping  of a supersonic beam of lithium atom used for atom interferometry. Because of the narrow hyperfine structure of the $^2$P$_{3/2}$ state, we have chosen to pump the atoms with the D1-line and the selected pumping scheme collects all the ground state population in a dark state, the $F=2,m_F= +2$ (or $-2$) sublevel. We developed a simplified model of the pumping process which predicts a high pumping efficiency, even if the polarization of the laser beams is not perfect. With previous theoretical and experimental works, we verified that the effects due to atom density, namely radiation trapping and spin-exchange collisions, should be negligible in our experiment.

Our atom interferometer requires an atomic beam with a very high transverse collimation, so that only a narrow transverse velocity class need to be pumped: this is an advantage because the laser beams used for the pumping does not require as much power as they would have needed for a broader class of transverse velocity. However,the weakness of the collimated beam makes it difficult to test the atom distribution over the ground state sublevels by the usual laser induced fluorescence technique so we characterized this distribution by two other techniques:

\begin{itemize}

\item following Gould \textit{et al.} \cite{GouldPRA87}, we have used a laser deflection experiment. This experiment has clearly shown that a large fraction of the population is collected in the aimed sublevel but because of a limited signal to noise ratio and insufficient resolution of the lines, we have not been able to put a precise upper limit on the residual populations of the other sublevels;

\item a magnetic dephasing experiment: as the Zeeman phase shift is a function of the sublevels, the visibility and phase of the interferometer signals are sensitive to the distribution of the atomic population over the $F,m_F$ sublevels. In the range studied here, the fringe phase is not very sensitive to a weak population distributed over the sublevels other than the $F=2, m_F=+2$ (or $-2$) but the fringe visibility is very sensitive to the population distribution.
\end{itemize}

With the second technique, we estimated the population in the aimed sublevel to be better than $95\pm 5$\%. The small variations of this efficiency with the sign of $m_F= \pm 2$ or with the mean velocity of the atomic beam are not well understood. This performance is close to the efficiency achieved by B.P. Masterson \textit{et al.} \cite{MastersonPRA93} with a cesium beam ($95$\% of the population in the aimed sublevel) or G.W. Schinn \textit{et al.} \cite{SchinnJOSAB91} with a sodium beam  ($97$\% of the population in the aimed sublevel). We used the optically pumped beam to perform a new measurement of the topological He-McKellar-Wilkens phase \cite{Gillot13} and optical pumping has considerably improved the experimental accuracy by eliminating systematic errors due to the average over $8$ sublevels \cite{LepoutrePRL12}.

We thank the laboratory technical staff for their help, G. Tr\'enec, A. Miffre and M. Jacquey for all the work they did on our atom interferometer. We thank CNRS INP, ANR (ANR-11-BS04-016-01 HIPATI) and R\'egion Midi-Pyr\'en\'ees for support.

\end{document}